\documentclass[aps,prl,twocolumn,superscriptaddress,showpacs]{revtex4}
\usepackage{amssymb}
\usepackage{graphicx}

\begin{document}

\title{Resistance Noise in Spin Valves}
\author{J\o rn Foros}
\affiliation{Department of Physics, Norwegian University of Science and Technology, 7491
Trondheim, Norway}
\author{Arne Brataas}
\affiliation{Department of Physics, Norwegian University of Science and Technology, 7491
Trondheim, Norway}
\author{Gerrit E. W. Bauer}
\affiliation{Kavli Institute of NanoScience, Delft University of Technology, 2628 CJ
Delft, The Netherlands}
\author{Yaroslav Tserkovnyak}
\affiliation{Lyman Laboratory of Physics, Harvard University, Cambridge, Massachusetts
02138, USA}
\date{\today}

\begin{abstract}
Fluctuations of the magnetization in spin valves are shown to cause
resistance noise that strongly depends on the magnetic configuration. Due to the applied
external field and the dynamic exchange interaction through the normal metal spacer,
the electrical noise level of the antiparallel configuration can
exceed that of the parallel one by an order of magnitude, in agreement with recent
experimental results.
\end{abstract}

\pacs{72.70.+m, 72.25.Mk, 75.75.+a}
\maketitle

The dynamics of nanoscale spin valve pillars in which electric currents are flowing
perpendicular to the interface planes (CPP) attract much interest \cite%
{Kiselev,Rippard,Covington}. The giant magnetoresistance (GMR) of such
pillars of ferromagnetic
films separated by normal metals, makes them attractive as future read heads in magnetic
hard-disk drives. However, CPP-GMR heads might be flawed by
noise. Covington \textit{et al.} \cite{Covington} found
enhanced low-frequency resistance noise in CPP spin valves. They ascribed this
to the spin-transfer torque, \textit{i.e.}, the torque exerted by a spin
polarized current on the magnetizations of the ferromagnetic layers \cite%
{Slonczewski1,Berger}. Rebei and Simionato \cite{Rebei} on the other hand, favored
micromagnetic disorder as an explanation. More recently, electrical noise
measurements have been carried out on CPP nanopillar multilayers with up to $%
15$ magnetic layers \cite{Covington2}. The noise\ power was found to be
reduced by more than an order of magnitude by aligning the magnetizations from
antiparallel to parallel in an external magnetic field.

The noise properties of small metallic structures pose challenges for
theoretical physics \cite{Blanter-review} to which ferromagnetism adds a
novel dimension \cite{Tserkovnyak-prb2001,FNF-SN,forosPRL}. The thermal
fluctuations of single domain magnetic clusters have been described already
50 years ago by Brown \cite{Brown}. Recently it has been shown that by contacting
the magnet with a conducting
environment, the magnetization fluctuations are enhanced compared to the
bulk value \cite{forosPRL}. Spin valves offer an opportunity to detect the
enhanced magnetization noise electrically by the GMR effect, but the new
degree of freedom of the detector magnetization complicates the picture in a
nontrivial way. The better understanding of the noise properties of CPP
nanopillar spin valves\ reported in the present Letter should therefore be
of interest for basic physics as well as applications.

In spin valves, two sources of thermal noise must be taken into account; direct
agitation of the magnetizations due to intrinsic processes \cite{Brown}, and thermal spin
current fluctuations outside the ferromagnets that affect the magnetizations by means of the
spin-transfer torque \cite{forosPRL}. Assuming low external bias, spin current shot-noise
(due to the discreteness of spin angular momentum) can be  
disregarded. We show in this Letter that the resulting
resistance noise strongly depends on the magnetic configuration. We find that
when the ferromagnets are ordered antiparallel, the noise level can indeed be an order of
magnitude higher than when they are parallel. Our results thus offer an explanation of
the experimental results of Covington \textit{et al.} \cite{Covington2}.

Resistance noise is defined 
\begin{equation}
S(t-t^{\prime })=\langle \Delta R(t)\Delta R(t^{\prime })\rangle ,
\label{Rnoise}
\end{equation}%
where $\Delta R(t)=R(t)-\langle R\rangle $ is the fluctuation of the
resistance at time $t$ from its time-averaged value. We consider a spin
valve with two ferromagnetic films with magnetizations $\mathbf{m}_{1}(t)$
and $\mathbf{m}_{2}(t)$ separated by a normal metal. At non-zero
temperatures both magnetizations have fluctuating parts $\delta \mathbf{m}%
_{1}(t)=\mathbf{m}_{1}(t)-\langle \mathbf{m}_{1}\rangle $ and $\delta 
\mathbf{m}_{2}(t)=\mathbf{m}_{2}(t)-\langle \mathbf{m}_{2}\rangle$,
due to intrinsic plus spin current noise. The resistance 
$R(t)$ of the spin valve depends on the angle $\theta $ between the
magnetizations. Close to collinear configurations, $R(t)$ can be expanded in
the small fluctuations as: 
\begin{equation}
R[\mathbf{m}_{1}(t)\cdot \mathbf{m}_{2}(t)]\approx R(\pm 1)\mp \frac{1}{2}%
[\delta \mathbf{m}^{\mp }(t)]^{2}\frac{\partial R}{\partial \cos \theta }%
\Big|_{P/AP},  \label{expandR}
\end{equation}%
where the upper/lower signs hold for the parallel (P)/antiparallel (AP)
orientation, $\delta \mathbf{m}^{\mp }(t)=\delta \mathbf{m}_{1}(t)\mp \delta 
\mathbf{m}_{2}(t),$ and the differential on the right hand side should be
evaluated for $\mathbf{m}_{1}\cdot \mathbf{m}_{2}=\cos \theta =1$ (P) or $%
\cos \theta =-1$ (AP). Eq. (\ref{expandR}) inserted in Eq. (\ref{Rnoise})
expresses the resistance noise in terms of the magnetization fluctuations $%
\delta \mathbf{m}^{\mp }(t)$. Only the difference between the magnetization
vectors $\delta \mathbf{m}^{-}(t)$ induces noise when the magnetizations are
parallel, whereas only the sum $\delta \mathbf{m}^{+}(t)$ contributes when
antiparallel. The fluctuations $\delta \mathbf{m}^{\mp }(t)$ are the
solutions of the stochastic equations of motion for the
magnetizations.

To start with, let us consider a single ferromagnetic film isolated from the
outside world. The magnetization dynamics in the macro-spin approximation is
given by the Landau-Lifshitz-Gilbert (LLG) equation: 
\begin{equation}
\frac{d\mathbf{m}}{dt}=-\gamma \mathbf{m}\times \lbrack \mathbf{H}_{\mathrm{%
eff}}+\mathbf{h}^{(0)}(t)]+\alpha _{0}\mathbf{m}\times \frac{d\mathbf{m}}{dt}%
,  \label{LLG}
\end{equation}%
where $\mathbf{m}$ is the unit magnetization vector, $\gamma $ the
gyromagnetic ratio, $\mathbf{H}_{\mathrm{eff}}$ the effective magnetic
field, $\alpha _{0}$ the Gilbert damping constant, and $\mathbf{h}^{(0)}(t)$
a time-dependent random field describing intrinsic thermal agitation. $\mathbf{h}%
^{(0)}(t)$ has zero average and a white noise correlation function that
satisfies the fluctuation-dissipation theorem (FDT) \cite{Brown}, 
\begin{equation}
\langle h_{i}^{(0)}(t)h_{j}^{(0)}(t^{\prime })\rangle =2k_{B}T\frac{\alpha
_{0}}{\gamma M_{s}\mathcal{V}}\delta _{ij}\delta (t-t^{\prime }).
\label{h0corr}
\end{equation}%
Here $i$ and $j$ are Cartesian components, $k_{B}T$ the thermal energy, $%
M_{s}$ the saturation magnetization, and $\mathcal{V}$ the volume of the
ferromagnet. When the ferromagnet is in contact with a conducting 
environment, two additional related effects may occur. First, when the magnetization
precesses, the ferromagnet emits spins into the neighboring conductors
(\textquotedblleft spin pumping\textquotedblright ) \cite{prl88}. In some cases, this
can be shown to be equivalent to an enhancement $\alpha ^{\prime }$ of the
Gilbert damping \cite{prl88}, $\alpha _{0}\rightarrow \alpha _{0}+\alpha
^{\prime }$. Secondly, a spin-polarized current (either due to spin pumping or external bias) can exert a torque on the
magnetization \cite{Berger,Slonczewski1,MyersKatine}, since its component
polarized transverse to the magnetization is absorbed at the interface \cite%
{prl84epjb22,Waintal,Stiles}. This spin-transfer torque may lead to
magnetization precession or complete reversal. In spin valves, spin pumping and corresponding spin-transfer
torque together couple the LLG-equations of the individual magnets
(\textquotedblleft dynamic exchange interaction\textquotedblright ) \cite%
{HeinrichBrataas,Tserkovnyakreview}.

Through the spin torque, thermal
spin current noise exerts a fluctuating torque on the magnetization vector.
As a result, the magnetization noise is increased as compared to the
intrinsic noise in an isolated ferromagnet. We have shown that for a single ferromagnet sandwiched by normal metals, the
enhancement of the noise is well described by a random field $\mathbf{h}^{\prime}(t)$ similar to, but statistically independent of, the
intrinsic random field $\mathbf{h}^{(0)}(t)$ \cite{forosPRL}. $\mathbf{h}^{\prime}(t)$
has correlation function equal to Eq. (\ref{h0corr}) but with the intrinsic
damping constant $\alpha _{0}$ replaced by $\alpha ^{\prime }$. Thus, the
magnetization noise is still governed by the FDT, with the total noise described by
the effective random field $\mathbf{h}(t)=\mathbf{h}^{(0)}(t)+\mathbf{h}^{\prime}(t)$ and the total dissipation by the enhanced Gilbert damping
constant $\alpha =\alpha _{0}+\alpha ^{\prime }$. The appearance of $\alpha
^{\prime }$ in this context identifies the thermal spin current noise as the
microscopic process that ensures validity of the FDT in the presence of spin
pumping \cite{forosPRL}.
\begin{figure}[tbp]
\includegraphics{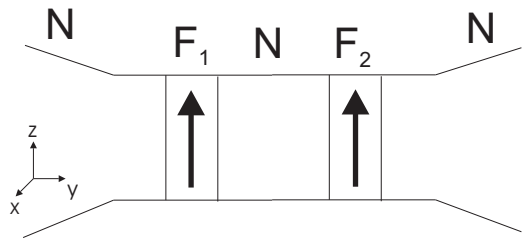}
\caption{A spin valve consists of two ferromagnetic thin films $F_{1}$ and $%
F_{2}$ separated by a normal metal spacer $N$ and connected to normal metal
reservoirs. Applied voltages are so small that their effect on the dynamics may be disregarded.}
\label{spinvalve}
\end{figure}

In an unbiased spin valve with a normal metal spacer thicker than
the spin-flip diffusion length, the ferromagnets move independently of
each other as described by Eq. (\ref{LLG}) with $\alpha _{0},\mathbf{h}^{(0)}
$ replaced by$\ \alpha ,\mathbf{h}$. More interesting is a spin valve with a
thin spacer through which the ferromagnets are allowed to communicate. The dynamic
exchange coupling due to spin pumping \cite%
{HeinrichBrataas,Tserkovnyakreview} must then be taken into account, together with the
static exchange coupling mediated by electrons in the normal metal spacer, and the
static coupling due to the magnetic dipolar interaction. The
interlayer static exchange coupling and dipolar coupling can both be described by a
Heisenberg coupling $-J\mathbf{m}_{1}\cdot \mathbf{m}_{2}$, where $J$ is the
coupling strength, favoring parallel (antiparallel) alignment when $J>0$ ($%
J<0$)$.$ We consider a spin valve such as the one pictured in Fig. \ref%
{spinvalve} in the regime of significant static and dynamic couplings. Two
ferromagnetic thin films are separated by a normal metal spacer and
connected to normal metal reservoirs. The ferromagnets are thicker than the
magnetic coherence length, so that they perfectly absorb any incoming spin
current polarized transverse to the magnetization direction \cite%
{prl84epjb22,Waintal,Stiles}. Spin-flip processes in the middle normal metal
are disregarded, which is usually allowed for CPP spin valves. Here we focus on the situation in
which the externally applied currents or voltages are sufficiently small to
not affect the device dynamics. For simplicity,
we take the spin valve to be symmetric (\textit{i.e}., the
two ferromagnets have identical conductance parameters), and only consider
parallel and antiparallel magnetic configurations. Assuming that the
exchange coupling $J$ is negative, the antiparallel state is the ground
state without applied external fields, while the parallel state is achieved
by applying a sufficiently strong external magnetic field forcing the
magnetizations to align.

Because of the static and dynamic coupling between the
ferromagnets, the resistance noise is shown below to vary substantially with
the magnetic configuration. The first step is to determine the total noise fields $\mathbf{h}_{k}(t)=%
\mathbf{h}_{k}^{(0)}(t)+\mathbf{h}_{k}^{\prime}(t)$ for ferromagnet $%
F_{1}$ ($k=1$) and ferromagnet $F_{2}$ ($k=2$). The intrinsic fields $\mathbf{h}%
_{k}^{(0)}(t)$ are given by Eq. (\ref{h0corr}), whereas
the spin current-induced fields $\mathbf{h}_{k}^{\prime}(t)$ can be calculated using
magnetoelectronic circuit theory \cite{prl84epjb22} and the results of Ref. \cite{forosPRL}.
Requiring conservation of charge and spin in the normal metal spacer \cite{Tserkovnyak-prb2001}, and taking
into account thermal fluctuations of the distribution function in the same spacer \cite{Tserkovnyak-prb2001},
the results are as follows \cite{foros}: For both the parallel and antiparallel configurations, 
the spin current-induced noise fields $\mathbf{h}_{1}^{\prime}(t)$ and $\mathbf{h}_{2}^{\prime}(t)$ are given by 
\begin{equation}
\langle h_{k,i}^{\prime}(t)h_{k,j}^{\prime}(t^{\prime
})\rangle =2k_{B}T\frac{\alpha _{\mathrm{sv}}^{\prime }}{\gamma M_{s}%
\mathcal{V}}\delta _{ij}\delta (t-t^{\prime }).  \label{hthcorr}
\end{equation}%
Here $k=1,2$, and $i$ and $j$ label axes perpendicular to the magnetization
direction. Furthermore, $\mathbf{h}_{1}^{\prime}(t)$ and $\mathbf{h}_{2}^{\prime}(t)$ are not
statistically independent: 
\begin{equation}
\langle h_{1,i}^{\prime}(t)h_{2,i}^{\prime}(t^{\prime
})\rangle =-\langle h_{1,i}^{\prime}(t)h_{1,i}^{\prime}(t^{\prime })\rangle   \label{hthcrosscorr}
\end{equation}
due to current conservation. As dictated by the FDT, $\alpha _{\mathrm{sv}}^{\prime }$ is
the spin-pumping enhancement of the Gilbert damping in each of the
ferromagnets of the spin valve (sv), given by  $\alpha _{\mathrm{sv}%
}^{\prime }=(\gamma \hbar \mathrm{Re}g^{\uparrow \downarrow })/(8\pi
M_{s}\mathcal{V})$ \cite{Tserkovnyakreview}. Here $g^{\uparrow \downarrow }$
is the dimensionless interface spin-mixing conductance (of which we have
disregarded a small imaginary part) \cite{prl84epjb22}. 
For simplicity, spin pumping and spin current fluctuations into
the outer normal metal reservoirs have been disregarded, but can, since 
the reservoirs are perfect spin sinks, be included simply by increasing
the intrinsic Gilbert damping constant $\alpha_0$.

We can now calculate the resistance noise, which is given by Eq. (\ref%
{expandR}) inserted in Eq. (\ref{Rnoise}). Assuming that the fluctuations of
the magnetization vectors are Gaussian distributed \cite{Brown}, we can
employ Wick's theorem \cite{Wick} and obtain 
\begin{equation}
S_{P/AP}(t-t^{\prime })=\frac{1}{2}\left( \frac{\partial R}{\partial (%
\mathrm{{cos}\theta )}}\right) _{P/AP}^{2}\sum_{i,j}S_{m_{i}^{\mp
}m_{j}^{\mp }}^{2}(t-t^{\prime }),  \label{afterWick}
\end{equation}%
for the resistance noise when the magnetizations are parallel (P, upper
sign) or antiparallel (AP, lower sign). Here $S_{m_{i}^{-}m_{j}^{-}}(t-t^{%
\prime })=\langle \delta {m}_{i}^{-}(t)\delta {m}_{j}^{-}(t^{\prime
})\rangle $, $S_{m_{i}^{+}m_{j}^{+}}(t-t^{\prime })=\langle \delta {m}%
_{i}^{+}(t)\delta {m}_{j}^{+}(t^{\prime })\rangle $, and the summation is
over all Cartesian components. $S_{m_{i}^{-}m_{j}^{-}}(t-t^{\prime })$ and $%
S_{m_{i}^{+}m_{j}^{+}}(t-t^{\prime })$ should be calculated from the LLG
equation, Eq. (\ref{LLG}), augmented to include the thermal spin current
noise, spin pumping, spin torque, and static exchange/dipolar coupling \cite%
{Tserkovnyakreview}: 
\begin{eqnarray}
\frac{d\mathbf{m}_{k}}{dt} &=&-\mathbf{m}_{k}\times \lbrack \omega _{0}\hat{\mathbf{z%
}}+\omega _{c}(\mathbf{m}_{k}\cdot \hat{\mathbf{x}})\hat{\mathbf{x}}+\omega _{x}\mathbf{m}%
_{l}+\gamma \mathbf{h}_{k}(t)]  \nonumber \\
&&+(\alpha _{0}+\alpha _{\mathrm{sv}}^{\prime })\mathbf{m}_{k}\times \frac{d%
\mathbf{m}_{k}}{dt}-\alpha _{\mathrm{sv}}^{\prime }\mathbf{m}_{l}\times 
\frac{d\mathbf{m}_{l}}{dt},  \label{LLGexpanded}
\end{eqnarray}%
where $k,l=1,2$ denotes ferromagnets 1 or 2, $\omega _{0}\hat{\mathbf{z}}=\gamma 
\mathbf{H}_{0}$ is an external field applied along the $z$-axis, $\omega
_{x}=\gamma J/M_{s}d$ parametrizes the static coupling (where $d$ is the
thickness of the ferromagnets), and $\mathbf{h}_{k}(t)=\mathbf{h}_{k}^{(0)}(t)+%
\mathbf{h}_{k}^{\prime}(t)$ is the total noise field. We have included an in-plane anisotropy field $%
\omega _{c}(\mathbf{m}_{k}\cdot \hat{\mathbf{x}})\hat{\mathbf{x}}=\gamma \mathbf{H}_{c}$ along
the $x$-axis. Both ferromagnets are described by the
damping parameter $\alpha =\alpha _{0}+\alpha _{\mathrm{sv}}^{\prime }$%
. The anisotropy field and the negative exchange coupling ($\omega _{x}<0$%
) align the ferromagnets antiparallel along the $x$-axis when the external
field is turned off. Then $\mathbf{m}_{k}(t)\approx\pm \hat{\mathbf{x}}+\delta \mathbf{m}%
_{k}(t)$ for $k=1,2$, where $\delta \mathbf{m}_{k}\approx \delta m_{k,y}\hat{%
\mathbf{y}}+\delta m_{k,z}\hat{\mathbf{z}}$ are the transverse fluctuations induced by the
random noise fields. Linearizing the LLG equation in $\delta \mathbf{m}_{k}$
we can evaluate the magnetization noise $S_{m_{i}^{+}m_{j}^{+}}(t-t^{\prime
})$ using Eqs. (\ref{h0corr}), (\ref{hthcorr}) and (\ref{hthcrosscorr}),
and find the resistance noise from Eq. (\ref{afterWick}). When turned on, a large external field
enforces a parallel magnetic configuration. Disregarding a sufficiently weak
anisotropy field, $\mathbf{m}_{k}(t)\approx \hat{\mathbf{z}}+\delta \mathbf{m}_{k}(t)$
for $k=1,2$, where $\delta \mathbf{m}_{k}\approx \delta m_{k,x}\hat{\mathbf{x}}%
+\delta m_{k,y}\hat{\mathbf{y}}$. This may be used to find $%
S_{m_{i}^{-}m_{j}^{-}}(t-t^{\prime })$ and subsequently $S_{P}(t-t^{\prime })
$.

The zero-frequency resistance noise $S_{P/AP}(\omega^{\prime} =0)=\int d(t-t^{\prime
})\langle \Delta R(t)\Delta R(t^{\prime })\rangle _{P/AP}$ thus becomes 
\begin{equation}
S_{P/AP}(0)=\frac{2}{\pi }\left( \frac{2\gamma k_{B}T}{M_{s}\mathcal{V}}%
\right) ^{2}\left( \frac{\partial R}{\partial \cos \theta }\right)
_{P/AP}^{2}\int d\omega X_{P/AP}
\label{zeroFreqRestNoise}
\end{equation}%
where 
\begin{equation}
X_{P}=\frac{[\omega ^{2}+(\omega _{t}-\omega _{c})^{2}]^{2}+(\omega
^{2}+\omega _{t}^{2})^{2}+2\omega ^{2}(2\omega _{t}-\omega _{c})^{2}}{%
2\alpha _{t}^{-2}([\omega ^{2}-\omega _{t}(\omega _{t}-\omega
_{c})]^{2}+\omega ^{2}\alpha _{t}^{2}[2\omega _{t}-\omega _{c}]^{2})^{2}}
\label{XP}
\end{equation}%
for the parallel and 
\begin{equation}
X_{AP}=\left( \frac{\omega ^{2}\alpha _{t}+\omega _{c}^{2}\alpha _{0}}{%
[\omega ^{2}+\omega _{c}(2\omega _{x}-\omega _{c})]^{2}+4\omega ^{2}[\omega
_{x}\alpha _{0}-\omega _{c}\alpha ]^{2}}\right) ^{2}
\label{XAP}
\end{equation}%
for the antiparallel configuration. Due to
the fact that the resistance noise goes as square of the magnetization noise (see Eq. (\ref{afterWick})),
there is an integration over frequency (coming from a Fourier transform) in the above expression.
The low-frequency resistance noise is hence not just given by the low-frequency magnetization noise.
$\omega _{t}=\omega_{0}+2\omega _{x}$ and $\alpha _{t}=\alpha _{0}+2\alpha ^{\prime }$ (note
the difference with $\alpha =\alpha _{0}+\alpha ^{\prime }$) correspond to the frequency and damping of the antisymmetric mode $\delta \mathbf{m}^{-}(t)$ in  
the P configuration \cite{Tserkovnyakreview}. We have set the external field to zero for the 
antiparallel configuration, and have assumed the damping to be small, $\alpha ^{2}\ll 1$, a condition which is
well obeyed by most ferromagnets \cite{Tserkovnyakreview}%
. The differential $\partial R/\partial \cos \theta $ can be calculated
using magnetoelectronic circuit theory \cite{prl84epjb22,foros}, and shown
to depend weakly on the magnetic configuration \cite{foros}. For simplicity, we approximate it in the following
to be a constant. The ratio $S_{AP}/S_{P}$ of the noise powers as a function
of the strength $-J$ of the static exchange coupling is shown in Fig. \ref%
{NoiseRatioMaple}, for some values of the applied external field in the parallel
configuration. Quite reasonably, the noise ratio increases with
increasing external field, since this field stabilizes the P configuration.
On the other hand, the noise ratio decreases with increasing coupling strength, because
the coupling stabilizes the AP configuration while destabilizing the P
configuration. 
\begin{figure}[tbp]
\includegraphics{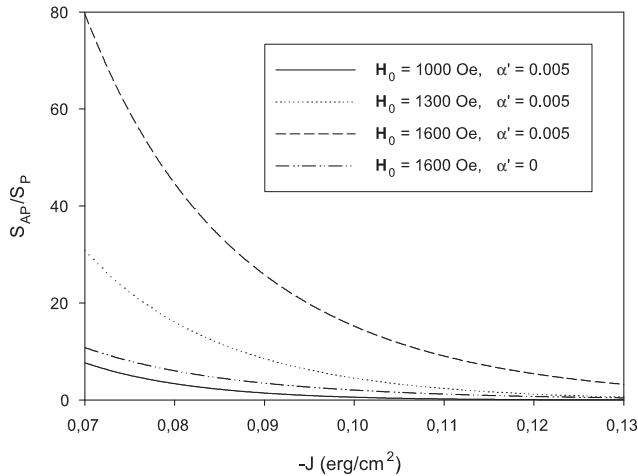}
\caption{The ratio $S_{AP}/S_{P}$ of the noise powers as a function of the
strength $\protect -J$ of the exchange coupling, for
some values of the applied external field in the parallel configuration (in the antiparallel configuration,
the external field is zero). The damping has been
set to $\protect\alpha _{0}=0.01$ and the anisotropy field to $\protect\omega_c/\protect\gamma=10$ Oe.}
\label{NoiseRatioMaple}
\end{figure}

Fig. \ref{NoiseRatioMaple} also emphasizes
the importance of including spin pumping and spin transfer torque. If
disregarded,\textit{\ i.e.}, $\alpha ^{\prime }=0$, the ratio $S_{AP}/S_{P}$
is substantially smaller. To understand this, consider the derivation of, and the expressions for $S_P$ and $S_{AP}$:
The noise $S_P$ is caused by the antisymmetric mode $%
\delta \mathbf{m}^{-}(t)=\delta \mathbf{m}_{1}(t)-\delta \mathbf{m}_{2}(t)$. This mode is subject to large
fluctuations as well as strongly damped by $\alpha _{t}=\alpha _{0}+2\alpha ^{\prime }$ \cite{Tserkovnyakreview},
in accordance with the FDT, and as can be seen from Eq. (\ref{XP}).
The noise $S_{AP}$ in the AP configuration on the other hand, is caused by
the mode $\delta \mathbf{m}^{+}(t)$, which is not as strongly damped.
At a first look, it may therefore seem like  
the FDT should imply larger resistance noise in the P configuration as compared to the AP.
But remember that the resistance noise goes as square of the magnetization noise,
resulting in the integration over frequency in Eq. (\ref{zeroFreqRestNoise}).
It is this fact, together with the difference in damping of the P and AP modes
(as appearing in the denominators of Eqs. (\ref{XP}) and (\ref{XAP})),
that is responsible for the suppression of the
resistance noise in the P configuration as compared to the AP.

We conclude from Fig. \ref{NoiseRatioMaple} that the noise strongly depends on the magnetic
configuration, and on the strength of the applied external field in the P configuration.
The noise level can be much higher in the antiparallel configuration than in the parallel, in agreement
with the experimental results of Covington \textit{et al.} \cite{Covington2} on nearly cylindrical multilayer pillars.
In these experiments the magnetizations were aligned parallel when the external magnetic field reached about 1500 Oe.
It must be noted that whereas we treated spin valves with two ferromagnetic films,
Covington \textit{et al.} dealt with multilayers of $4-15$ magnetic films. It is
likely that the difference between the noise properties of bilayers and
multilayers is small, as the only local structural difference is the number
of neighboring ferromagnets. This assertion is supported by the experiments by Covington \textit{et
al.} that did not reveal strong differences for nanopillars ranging from $%
4-15$ layers.

We thank Mark Covington for sharing his results prior to publication,
and Hans J. Skadsem for discussions. This work was supported in part by the
Research Council of Norway, NANOMAT Grants No. 158518/143
and 158547/431, the EU Commission FP6 NMP-3 project 505587-1
\textquotedblleft SFINX\textquotedblright , and the FOM.


\begin{thebibliography}{21}
\expandafter\ifx\csname natexlab\endcsname\relax\def\natexlab#1{#1}\fi
\expandafter\ifx\csname bibnamefont\endcsname\relax
  \def\bibnamefont#1{#1}\fi
\expandafter\ifx\csname bibfnamefont\endcsname\relax
  \def\bibfnamefont#1{#1}\fi
\expandafter\ifx\csname citenamefont\endcsname\relax
  \def\citenamefont#1{#1}\fi
\expandafter\ifx\csname url\endcsname\relax
  \def\url#1{\texttt{#1}}\fi
\expandafter\ifx\csname urlprefix\endcsname\relax\def\urlprefix{URL }\fi
\providecommand{\bibinfo}[2]{#2}
\providecommand{\eprint}[2][]{\url{#2}}

\bibitem[{\citenamefont{Kiselev et~al.}(2003)\citenamefont{Kiselev, Sankey,
  Krivorotov, Emley, Schoelkopf, Buhrman, and Ralph}}]{Kiselev}
\bibinfo{author}{\bibfnamefont{S.~I.} \bibnamefont{Kiselev}},
  \bibinfo{author}{\bibfnamefont{J.~C.} \bibnamefont{Sankey}},
  \bibinfo{author}{\bibfnamefont{I.~N.} \bibnamefont{Krivorotov}},
  \bibinfo{author}{\bibfnamefont{N.~C.} \bibnamefont{Emley}},
  \bibinfo{author}{\bibfnamefont{R.~J.} \bibnamefont{Schoelkopf}},
  \bibinfo{author}{\bibfnamefont{R.~A.} \bibnamefont{Buhrman}},
  \bibnamefont{and} \bibinfo{author}{\bibfnamefont{D.~C.} \bibnamefont{Ralph}},
  \bibinfo{journal}{Nature} \textbf{\bibinfo{volume}{425}},
  \bibinfo{pages}{380} (\bibinfo{year}{2003}).

\bibitem[{\citenamefont{Rippard et~al.}(2004)\citenamefont{Rippard, Pufall,
  Kaka, Russek, and Silva}}]{Rippard}
\bibinfo{author}{\bibfnamefont{W.~H.} \bibnamefont{Rippard}},
  \bibinfo{author}{\bibfnamefont{M.~R.} \bibnamefont{Pufall}},
  \bibinfo{author}{\bibfnamefont{S.}~\bibnamefont{Kaka}},
  \bibinfo{author}{\bibfnamefont{S.~E.} \bibnamefont{Russek}},
  \bibnamefont{and} \bibinfo{author}{\bibfnamefont{T.~J.} \bibnamefont{Silva}},
  \bibinfo{journal}{Phys. Rev. Lett.} \textbf{\bibinfo{volume}{92}},
  \bibinfo{pages}{027201} (\bibinfo{year}{2004}).

\bibitem[{\citenamefont{Covington et~al.}(2004)\citenamefont{Covington,
  AlHajDarwish, Ding, Gokemeijer, and Seigler}}]{Covington}
\bibinfo{author}{\bibfnamefont{M.}~\bibnamefont{Covington}},
  \bibinfo{author}{\bibfnamefont{M.}~\bibnamefont{AlHajDarwish}},
  \bibinfo{author}{\bibfnamefont{Y.}~\bibnamefont{Ding}},
  \bibinfo{author}{\bibfnamefont{N.~J.} \bibnamefont{Gokemeijer}},
  \bibnamefont{and} \bibinfo{author}{\bibfnamefont{M.~A.}
  \bibnamefont{Seigler}}, \bibinfo{journal}{Phys. Rev. B}
  \textbf{\bibinfo{volume}{69}}, \bibinfo{pages}{184406}
  (\bibinfo{year}{2004}).

\bibitem[{\citenamefont{Slonczewski}(1996)}]{Slonczewski1}
\bibinfo{author}{\bibfnamefont{J.~C.} \bibnamefont{Slonczewski}},
  \bibinfo{journal}{J. Magn. Magn. Mater.} \textbf{\bibinfo{volume}{159}},
  \bibinfo{pages}{L1} (\bibinfo{year}{1996}).

\bibitem[{\citenamefont{Berger}(1996)}]{Berger}
\bibinfo{author}{\bibfnamefont{L.}~\bibnamefont{Berger}},
  \bibinfo{journal}{Phys. Rev. B} \textbf{\bibinfo{volume}{54}},
  \bibinfo{pages}{9353} (\bibinfo{year}{1996}).

\bibitem[{\citenamefont{Rebei and Simionato}(2005)}]{Rebei}
\bibinfo{author}{\bibfnamefont{A.}~\bibnamefont{Rebei}} \bibnamefont{and}
  \bibinfo{author}{\bibfnamefont{M.}~\bibnamefont{Simionato}},
  \bibinfo{journal}{Phys. Rev. B} \textbf{\bibinfo{volume}{71}},
  \bibinfo{pages}{174415} (\bibinfo{year}{2005}).

\bibitem[{Cov()}]{Covington2}
\bibinfo{note}{M. Covington, to be published}.

\bibitem[{\citenamefont{Blanter and B{\"{u}}ttiker}(2000)}]{Blanter-review}
\bibinfo{author}{\bibfnamefont{Y.~M.} \bibnamefont{Blanter}} \bibnamefont{and}
  \bibinfo{author}{\bibfnamefont{M.}~\bibnamefont{B{\"{u}}ttiker}},
  \bibinfo{journal}{Phys. Rep.} \textbf{\bibinfo{volume}{336}},
  \bibinfo{pages}{1} (\bibinfo{year}{2000}).

\bibitem[{\citenamefont{Tserkovnyak and Brataas}(2001)}]{Tserkovnyak-prb2001}
\bibinfo{author}{\bibfnamefont{Y.}~\bibnamefont{Tserkovnyak}} \bibnamefont{and}
  \bibinfo{author}{\bibfnamefont{A.}~\bibnamefont{Brataas}},
  \bibinfo{journal}{Phys. Rev. B} \textbf{\bibinfo{volume}{64}},
  \bibinfo{pages}{214402} (\bibinfo{year}{2001}).

\bibitem[{FNF()}]{FNF-SN}
\bibinfo{note}{E. G. Mishchenko, Phys. Rev. B \textbf{68}, 100409(R) (2003); D.
  S\'{a}nchez, R. L\'{o}pez, P. Samuelsson, and M. B{\"{u}}ttiker, Phys. Rev. B
  \textbf{68}, 214501 (2003); A. Lamacraft, Phys. Rev. B \textbf{69}, 081301(R)
  (2004); W. Belzig and M. Zareyan, Phys. Rev. B \textbf{69}, 140407(R)
  (2004)}.

\bibitem[{\citenamefont{Foros et~al.}(2005)\citenamefont{Foros, Brataas,
  Tserkovnyak, and Bauer}}]{forosPRL}
\bibinfo{author}{\bibfnamefont{J.}~\bibnamefont{Foros}},
  \bibinfo{author}{\bibfnamefont{A.}~\bibnamefont{Brataas}},
  \bibinfo{author}{\bibfnamefont{Y.}~\bibnamefont{Tserkovnyak}},
  \bibnamefont{and} \bibinfo{author}{\bibfnamefont{G.~E.~W.}
  \bibnamefont{Bauer}}, \bibinfo{journal}{Phys. Rev. Lett.}
  \textbf{\bibinfo{volume}{95}}, \bibinfo{pages}{016601}
  (\bibinfo{year}{2005}).

\bibitem[{\citenamefont{Brown}(1963)}]{Brown}
\bibinfo{author}{\bibfnamefont{W.~F.} \bibnamefont{Brown}},
  \bibinfo{journal}{Phys. Rev.} \textbf{\bibinfo{volume}{130}},
  \bibinfo{pages}{1677} (\bibinfo{year}{1963}).

\bibitem[{\citenamefont{Tserkovnyak et~al.}(2002)\citenamefont{Tserkovnyak,
  Brataas, and Bauer}}]{prl88}
\bibinfo{author}{\bibfnamefont{Y.}~\bibnamefont{Tserkovnyak}},
  \bibinfo{author}{\bibfnamefont{A.}~\bibnamefont{Brataas}}, \bibnamefont{and}
  \bibinfo{author}{\bibfnamefont{G.~E.~W.} \bibnamefont{Bauer}},
  \bibinfo{journal}{Phys. Rev. Lett.} \textbf{\bibinfo{volume}{88}},
  \bibinfo{pages}{117601} (\bibinfo{year}{2002}).

\bibitem[{Mye()}]{MyersKatine}
\bibinfo{note}{E. B. Myers, D. C. Ralph, J. A. Katine, R. N. Louie, and R. A.
  Buhrman, Science \textbf{285}, 867 (1999); J. A. Katine, F. J. Albert, R. A.
  Buhrman, E. B. Myers, and D. C. Ralph, Phys. Rev. Lett. \textbf{84}, 3149
  (2000)}.

\bibitem[{prl()}]{prl84epjb22}
\bibinfo{note}{A. Brataas, Yu. V. Nazarov, and G. E. W. Bauer, Phys. Rev. Lett.
  \textbf{84}, 2481 (2000); Eur. Phys. J. B \textbf{22}, 99 (2001); A. Brataas,
  G. E. W. Bauer, and P. J. Kelly, Phys. Rep. \textbf{427}, 157 (2006)}.

\bibitem[{\citenamefont{Waintal et~al.}(2000)\citenamefont{Waintal, Myers,
  Brouwer, and Ralph}}]{Waintal}
\bibinfo{author}{\bibfnamefont{X.}~\bibnamefont{Waintal}},
  \bibinfo{author}{\bibfnamefont{E.~B.} \bibnamefont{Myers}},
  \bibinfo{author}{\bibfnamefont{P.~W.} \bibnamefont{Brouwer}},
  \bibnamefont{and} \bibinfo{author}{\bibfnamefont{D.~C.} \bibnamefont{Ralph}},
  \bibinfo{journal}{Phys. Rev. B} \textbf{\bibinfo{volume}{62}},
  \bibinfo{pages}{12317} (\bibinfo{year}{2000}).

\bibitem[{\citenamefont{Stiles and Zangwill}(2002)}]{Stiles}
\bibinfo{author}{\bibfnamefont{M.~D.} \bibnamefont{Stiles}} \bibnamefont{and}
  \bibinfo{author}{\bibfnamefont{A.}~\bibnamefont{Zangwill}},
  \bibinfo{journal}{Phys. Rev. B} \textbf{\bibinfo{volume}{66}},
  \bibinfo{pages}{014407} (\bibinfo{year}{2002}).

\bibitem[{\citenamefont{Heinrich et~al.}(2003)\citenamefont{Heinrich,
  Tserkovnyak, Woltersdorf, Brataas, Urban, and Bauer}}]{HeinrichBrataas}
\bibinfo{author}{\bibfnamefont{B.}~\bibnamefont{Heinrich}},
  \bibinfo{author}{\bibfnamefont{Y.}~\bibnamefont{Tserkovnyak}},
  \bibinfo{author}{\bibfnamefont{G.}~\bibnamefont{Woltersdorf}},
  \bibinfo{author}{\bibfnamefont{A.}~\bibnamefont{Brataas}},
  \bibinfo{author}{\bibfnamefont{R.}~\bibnamefont{Urban}}, \bibnamefont{and}
  \bibinfo{author}{\bibfnamefont{G.~E.~W.} \bibnamefont{Bauer}},
  \bibinfo{journal}{Phys. Rev. Lett.} \textbf{\bibinfo{volume}{90}},
  \bibinfo{pages}{187601} (\bibinfo{year}{2003}).

\bibitem[{\citenamefont{Tserkovnyak et~al.}(2005)\citenamefont{Tserkovnyak,
  Brataas, Bauer, and Halperin}}]{Tserkovnyakreview}
\bibinfo{author}{\bibfnamefont{Y.}~\bibnamefont{Tserkovnyak}},
  \bibinfo{author}{\bibfnamefont{A.}~\bibnamefont{Brataas}},
  \bibinfo{author}{\bibfnamefont{G.~E.~W.} \bibnamefont{Bauer}},
  \bibnamefont{and} \bibinfo{author}{\bibfnamefont{B.~I.}
  \bibnamefont{Halperin}}, \bibinfo{journal}{Rev. Mod. Phys.}
  \textbf{\bibinfo{volume}{77}}, \bibinfo{pages}{1375} (\bibinfo{year}{2005}).

\bibitem[{\citenamefont{Foros et~al.}()\citenamefont{Foros, Brataas, Bauer, and
  Tserkovnyak}}]{foros}
\bibinfo{author}{\bibfnamefont{J.}~\bibnamefont{Foros}},
  \bibinfo{author}{\bibfnamefont{A.}~\bibnamefont{Brataas}},
  \bibinfo{author}{\bibfnamefont{G.~E.~W.} \bibnamefont{Bauer}},
  \bibnamefont{and}
  \bibinfo{author}{\bibfnamefont{Y.}~\bibnamefont{Tserkovnyak}},
  \bibinfo{note}{to be published}.

\bibitem[{\citenamefont{Triantafyllopoulos}(2003)}]{Wick}
\bibinfo{author}{\bibfnamefont{K.}~\bibnamefont{Triantafyllopoulos}},
  \bibinfo{journal}{The Mathematical Scientist} \textbf{\bibinfo{volume}{28}},
  \bibinfo{pages}{125} (\bibinfo{year}{2003}).

\end{thebibliography}
\end{document}